\documentclass[prb,preprint,aps,superscriptaddress]{revtex4-1}
\usepackage{hyperref}
\usepackage{amsfonts,amssymb}
\usepackage{graphicx}
\usepackage{color,soul}
\usepackage{url}
\usepackage{amsmath}
\usepackage{dcolumn}
\usepackage{latexsym,bm}
\usepackage{tabularx}
\usepackage{enumerate}
\usepackage{mathtools}

\usepackage[T1]{fontenc}       

\begin{document}

\title{Theoretical Prediction of Two-Dimensional Functionalized MXene Nitrides as Topological Insulators}

\author{Yunye Liang}
\email{liangyunye@shnu.edu.cn}
\affiliation{Department of Physics, Shanghai Normal University, Shanghai 200234, China}

\author{Mohammad Khazaei}
\affiliation{Computational Materials Science Research Team, RIKEN Advanced Institute for Computational Science (AICS), Kobe, Hyogo 650-0047, Japan}

\author{Ahmad Ranjbar}
\affiliation{Computational Materials Science Research Team, RIKEN Advanced Institute for Computational Science (AICS), Kobe, Hyogo 650-0047, Japan}

\author{Masao Arai}
\affiliation{International Center for Materials Nanoarchitectonics, National Institute for Materials Science (NIMS), 1-1 Namiki, 
Tsukuba 305-0044, Ibaraki, Japan}

\author{Seiji Yunoki}
\affiliation{Computational Materials Science Research Team, RIKEN Advanced Institute for Computational Science (AICS), Kobe, Hyogo 650-0047, Japan}
\affiliation{Computational Condensed Matter Physics Laboratory, RIKEN, Wako, Saitama 351-0198, Japan}
\affiliation{Computational Quantum Matter Research Team, RIKEN Center for Emergent Matter Science (CEMS), Wako, Saitama 351-0198, Japan}

\author{Yoshiyuki Kawazoe}

\affiliation{New Industry Creation Hatchery Center, Tohoku University,
  Sendai, 980-8579, Japan}

\affiliation{Department of Physics and Nanotechnology, SRM University, Katankulathur, Chennai 603203,
Tamil Nadu, India}

\author{Hongming Weng}
\affiliation{Beijing National Laboratory for Condensed Matter Physics,
  and Institute of Physics, Chinese Academy of Sciences, Beijing
  100190, China}

\affiliation{Collaborative Innovation Center of Quantum Matter,
  Beijing, China}

\author{Zhong Fang}
\affiliation{Beijing National Laboratory for Condensed Matter Physics,
  and Institute of Physics, Chinese Academy of Sciences, Beijing
  100190, China}

\affiliation{Collaborative Innovation Center of Quantum Matter,
  Beijing, China}

\date{\today}

\begin{abstract}

Recently, two-dimensional (2D) transition metal carbides and nitrides, namely, MXenes have attracted lots of 
attention for electronic and energy storage applications. Due to a large spin-orbit coupling (SOC) and the existence 
of a Dirac-like band at the Fermi energy, it has been theoretically proposed that some of the MXenes will be 
topological insulators (TIs). Up to now, all of the predicted TI MXenes belong to transition metal carbides, 
whose transition metal atom is W, Mo or Cr. Here, on the basis of first-principles and $\mathbb{Z}_2$ index 
calculations, we demonstrate that some of the MXene nitrides can also be TIs. We find that Ti$_3$N$_2$F$_2$ is a 2D TI, whereas Zr$_3$N$_2$F$_2$ is a semimetal with nontrivial band topology and can be turned into a 2D TI when the lattice is stretched. We also find that the tensile strain can convert Hf$_3$N$_2$F$_2$ semiconductor into a 2D TI. Since Ti is one of the mostly used transition metal element in the synthesized MXenes, we expect that our prediction can advance the future application of MXenes as TI devices.
\end{abstract}

\maketitle

\section{Introduction} \label{introduction}

At the beginning, graphene was proposed as a two-dimensional (2D) topological insulator 
(TI)~\cite{C. L. Kane1, C. L. Kane2, Yao1,B. Andrei Bernevig1,M. Z. Hasan,X. L. Qi,Hongming Weng1}. 
Unfortunately, the band gap of graphene is too tiny to be measured experimentally. Later on, the observation 
of topological quantum states (TQSs) was realized in a 2D quantum well of 
HgTe/CdTe~\cite{B. Andrei Bernevig2,Markus Konig}. This  experimental achievement boosted the quick rising 
of the field of TIs. For example, the idea of band topology was extended  to three-dimensional 
(3D) systems and other symmetry  protected  TQSs~\cite{Liang Fu0,Liang Fu,J. E. Moore,Rahul Roy,C.-K. Chiu}. Recently, the band topology in metals,~\cite{TSM} including the topological Dirac semimetal~\cite{Zhijun Wang, Zhijun Wang2, Z. K. Liu, Z. K. Liu2}, Weyl semimetal~\cite{Xiangang Wan, Gang Xu, Hongming Weng2, B. Q. Lv, B. Q. Lv2, SMHuang, SYXu}, and node-line semimetal~\cite{A. A. Burkov, H. Weng, R. Yu, Y. Kim} has also been intensively studied. Many of materials with TQSs were firstly predicted by theoretical calculations before the experimental confirmations~\cite{Hongming Weng1, Haijun Zhang}. 2D TIs are particularly useful 
in applications because they are expected to host quantum spin Hall effect (QSHE) with one-dimensional 
helical edge states. The electrons in such edge states have opposite velocities in opposite spin channels, 
and thus the backscattering is prohibited as long as the perturbation does not break the time-reversal 
symmetry. Such helical edge states are expected to serve as ``two-lane highway'' for dissipationless 
electron transport, which promises great potential application 
in low-power and multi-functional spintronic devices.

Compared with the number of well characterized 3D TI materials, fewer 2D TIs have been experimentally discovered~\cite{Hongming Weng1,Yoichi Ando}. The quantum wells of HgTe/CdTe~\cite{Markus Konig} 
and InAs/GaSb~\cite{Ivan Knez} are among the well-known experimentally confirmed 2D TIs. 
The requirements of precise control of molecular beam epitaxy growth and operation at ultra-low 
temperatures make further studies hard for the possible applications. Different  methods  are used to find or 
create prospective  TIs, for examples, by tuning the band gaps through the  strength of spin-orbit  coupling (SOC) 
in graphene-like honeycomb lattice structures, such as the low-buckled silicene~\cite{Cheng-Cheng Liu}, 
chemically decorated single layer honeycomb lattice of Sn~\cite{Yong Xu}, Ge~\cite{Chen Si2}, Bi, or 
Sb ~\cite{Z. Song}, and bucked square lattice BiF~\cite{Wei Luo}, or by 
examining new 2D systems, which might be exfoliated from the 3D layered structural materials, such as 
ZrTe$_5$, HfTe$_5$~\cite{Hongming Weng3, Wu1}, Bi$_4$Br$_4$, and 
Bi$_4$I$_4$~\cite{Jin-Jian Zhou, Oleg}. Transition-metal 
dichalcogenide~\cite{Xiaofeng Qian, WTe21,WTe22,WTe23,WTe24} and square-octagon 
haeckelite~\cite{S. M. Nie,Y. Sun,Y. Ma} structure belong to the later category. Among these proposals, 
ZrTe$_5$, HfTe$_5$, Bi$_4$Br$_4$, Bi$_4$I$_4$, WTe$_2$ and Bismuthene on SiC substrate seem to be very promising. \cite{F. Reis} 


MXenes are a novel family of 2D transition metal 
carbonitrides~\cite{Michael Naguib,Michael Naguib2,Michael Naguib3} which have been recently obtained 
by the selective chemical etching of MAX phases --- M$_{n+1}$AX$_n$ ($n$=1-3), where M, A, and X are a 
transition metal, an element of group 13-14, and C or N, respectively~\cite{Michel W. Barsoum}. 
MXenes have some unique properties and applications~\cite{khazaei-review,babak}. 
As examples, it is predicted that MXenes can be used as energy storage devices such as batteries and 
electrochemical supercapacitors~\cite{M. R. Lukatskaya,O. Mashtalir2}. This is because a variety of cations 
with various charges and sizes, including Na$^+$, K$^+$, NH$_4^+$, Mg$^{2+}$, and Al$^{3+}$, 
can be easily intercalated into the exfoliated MXenes~\cite{M. R. Lukatskaya}.

The flexibility of MXenes make them prospective candidates of the wearable energy storage devices~\cite{J. R. Miller}. The bare surfaces of MXene sheets are chemically active and are usually terminated with atoms or chemical groups, depending on the synthesis process, usually fluorine (F), oxygen (O), or hydroxyl (OH)~\cite{Kristopher J. Harris, Mohammad Khazaei, Mohammad Khazaei2}. 
Up to now, many 2D MXenes such as Ti$_2$C, V$_2$C, Nb$_2$C, Mo$_2$C, Ti$_3$CN, Ti$_3$C$_2$, Ta$_4$C$_3$, and Ti$_3$C$_2$ have been experimentally produced~\cite{Michael Naguib2,Michael Naguib3,Michael Naguib4,meshkini2015}. More recently, double ordered transition metals MXenes, Mo$_2$TiC$_2$ and Mo$_2$Ti$_2$C$_3$, have also been synthesized~\cite{anasori2016_1,anasori2016_2}.

Various 2D MXenes have been predicted to be TIs. In this regard, we have reported that the functionalized MXenes with oxygen, M$_2$CO$_2$ (M=W, Mo, and Cr) are 2D TIs.~\cite{Hongming Weng4}  The results are robust against the use of different exchange-correlation functional approximations. The bulk band gap of W$_2$CO$_2$ is as large as 0.194 eV within generalized 
gradient approximation (GGA)~\cite{Georg Kresse,Georg Kresse2,John Perdew} and is enhanced to 0.472 eV 
within non-local Heyd-Scuseria-Ernzerhof (HSE06) hybrid functional~\cite{Jochen Heyd, Jochen Heyd2}. 
TIs can also be found in ordered double transition metals M$'_2$M$''$C$_2$ MXenes, where M$'$=Mo and W, and M$''$=Ti, Zr, and Hf~\cite{Mohammad Khazaei2016,Chen Si,Longhua Li}. A 2D TI with large band gap is crucial to realize the long-sought-for topological superconductivity and Majorana modes through proximity effect~\cite{X. L. Qi,Liang Fu2}. Inspired by these findings, we have studied a large number of 2D MXenes in order to obtain new TI candidates with different topologically nontrivial properties that may be easily synthesized~\cite{Michel W. Barsoum, Mohammad Khazaei3}. The possible variants of 2D MXenes are achieved by replacing O$^{2-}$ by F$^{-}$ or (OH)$^{-}$~\cite{Mohammad Khazaei,Mohammad Khazaei2, H. Fashandi}, substituting transition-metal M, varying C with N or B~\cite{Mohammad Khazaei3}, and tuning the number of layers $n$ in MXene M$_{n+1}$C$_n$~\cite{Mohammad Khazaei4,Wang Xuefeng}. Properly tailoring materials with a single or any combination of the above may lead to more and better 2D topologically insulating MXenes.

In this paper, by using first-principles calculations, we predict that functionalized MXenes Ti$_3$N$_2$F$_2$ 
and Zr$_3$N$_2$F$_2$ have nontrivial band topology: the former is a 2D TI and the latter is a semimetal which can be turned into a 2D TI by the tensile strain. We also predict that Hf$_3$N$_2$F$_2$ semiconductor can be converted into TI by the tensile strain. The stabilities of the structures are confirmed by phonon dispersion calculations. The band inversion, which is crucial to the nontrivial band topology, is found to occur among the different combinations of $d$-orbitals due to the crystal field effect. The results are robust against different exchange-correlation functional approximations such as the GGA and the HSE06 hybrid functional. 

The rest of this paper is organized as follows. First, the computational details are briefly summarized 
in Sec.~\ref{method}. The main results together with discussion are provided in 
Sec.~\ref{Results}. Finally, the paper is concluded in Sec.~\ref{sec:conclusion}.

\section{Computational Details} \label{method}

Calculations are performed by Vienna $ab~ initio$ simulation package (VASP), 
within the Perdew-Burke-Ernzerhof  (PBE) version of the GGA functional for exchange 
correlation~\cite{Georg Kresse,Georg Kresse2,John Perdew}. The SOC is taken into account self-consistently. 
The cut-off energy is 520 eV for plane wave expansion.  The optimization calculations are 
done with 12$\times$12$\times$1 k-point sampling grid. The crystal structures are fully relaxed 
until the residual forces on each atom become less than 0.001 eV/\AA. 
A vacuum space between layers is 40 \AA, in order to minimize the interactions between the layer with its 
periodic images. PHONOPY is employed to calculate the phonon dispersion~\cite{Atsushi Togo}. 
Considering the possible underestimation of band gap within the GGA, the non-local HSE06 hybrid functional~\cite{Jochen Heyd, Jochen Heyd2} is further applied to check the band topology. 
To explore the edge states, we apply the Green's function method~\cite{Hongming Weng1} based on the 
tight-binding model with the maximally localized Wannier functions (MLWFs)~\cite{Nicola Marzari,Ivo Souza} 
of $d$ orbitals of Ti and $p$ orbitals of N and F as the basis set. The MLWFs are generated by using the 
software package OpenMX~\cite{openmx, Hongming Weng5}.

\section{Results and Discussion} \label{Results}

\subsection{Atomic Structure}

The  bare  MXenes are chemically reactive. Hence, depending on the synthesis process, the transition metals 
on the surfaces of the MXenes are functionalized by a mixture of F, O, and OH. First-principles calculations show that the electronic properties of MXenes can be engineered by appropriate surface 
functionalization~\cite{khazaei-review,Mohammad Khazaei, Mohammad Khazaei2,Mohammad Khazaei5}. 
The MXenes with full surface functionalizations, namely, two chemical groups on both sides of the surfaces, 
are thermodynamically more favorable than those with the partial functionalizations~\cite{Mohammad Khazaei}. 
Figures~\ref{crystructure}(a) and \ref{crystructure}(b) show the crystal structure of functionalized MXene Ti$_3$N$_2$F$_2$, 
which consists of seven atomic layers with hexagonal lattice, including three Ti layers, two N layers, and 
two F layers. 

As shown in Fig.~\ref{crystructure}(b), the chemical  functional groups can occupy possibly three different sites, 
indicated by A, B, and T, on each surface. Therefore, there are six structurally possible configurations 
which should be considered, as listed in Table~\ref{stablestructure}. The atomic positions and lattice vectors for these six different configurations are fully optimized and the total energies of the optimized structures are summarized in Table~\ref{stablestructure}. The calculations reveal that the MXene with BB-type fluorine functionalization has the lowest energy. 

The space group of BB-type Ti$_3$N$_2$F$_2$ is \emph{P$\overline{3}$m1}, and there are two inequivalent types of Ti 
atoms: the inner Ti atomic layer and the outer Ti atomic layers. The Wyckoff position of the inner Ti atom is $1a$ 
and it is the origin of the primitive cell. The Wyckoff position of the outer Ti atoms is $2d$ and the corresponding 
coordinates are $(1/3,2/3,z)$ and $(1/3,2/3,\overline{z})$. 

Since there is no report of the successful experimental synthesis of 2D Ti$_3$N$_2$F$_2$ 
in the literatures yet, the phonon dispersion spectrum of the BB structure is calculated 
to confirm the structural stability. As shown in Fig.~\ref{crystructure}(c), all phonon frequencies are positive, 
indicating that the structure is dynamically stable. 


 \begin{table}[tb]
\caption{
The total energies (in eV per unit cell) for the six possible configurations of $M$$_3$N$_2$F$_2$, where $M$ 
is Ti, Zr, and Hf. In each unit cell, fluorine atoms on both surfaces are required for full surface 
saturations. T, A, and B indicate three different absorption positions, as indicated in Fig.~\ref{crystructure}(b).  
All calculations here are performed within GGA and without spin polarization. 
The energy of the most favorable structure (BB) is set to zero. 
}\label{stablestructure}
\begin{tabular*}{0.45\textwidth}{@{\extracolsep{\fill}}c|c|c|c}
 	\hline\hline
   sites of Fluorine &Ti$_3$N$_2$F$_2$&Zr$_3$N$_2$F$_2$&Hf$_3$N$_2$F$_2$    \\
    \hline
TT & 1.667&2.119&1.761 \\
 	\hline
AA & 0.093  &0.388&0.634 \\
 	\hline
TB & 1.025  &1.072&0.886 \\
 	\hline
TA  & 0.873 &1.256&1.226  \\
         \hline
BB  & 0  &0&0\\
         \hline
BA  &  0.090&0.234&0.379\\
 	\hline\hline
\end{tabular*}
\end{table}

\subsection{Electronic Structure of Ti$_3$N$_2$F$_2$}

Next, we examine the electronic structure of Ti$_3$N$_2$F$_2$ with the optimized atomic structure obtained 
above. As shown in Fig.~\ref{bands}(a), if the SOC is not considered, the lowest conduction band and the highest 
valence band touch each other at the $\Gamma$ point exactly at the Fermi energy. They are two-fold degenerate 
and characterized by the irreducible representation $E_g$. The PBE calculations with the SOC show 
in Fig.~\ref{bands}(b)  
that Ti$_3$N$_2$F$_2$ may be semimetal with slightly having compensated electron and hole Fermi pocket. 
However, as shown in Fig.~\ref{bands}(b), the more accurate calculations using HSE06 with the SOC show that 
Ti$_3$N$_2$F$_2$ is an insulator.

Given the fact that the primitive cell of Ti$_3$N$_2$F$_2$ possesses the inversion symmetry and following 
the method proposed by Fu and Kane~\cite{Liang Fu}, the topological invariant $\nu$ reads as 
\begin{equation}
\delta(k_i)=\prod_{n=1}^N\xi_{n} (k_i);   \quad\quad\quad 
(-1)^\nu=\prod_{i=1}^4\delta(k_i)
\label{eq1}
\end{equation}
where $\xi_n(k_i)$ is $+1$ ($-1$) for even (odd) parity of the $n$th occupied band 
(not including the Kramers degenerate partner) at 
time reversal invariant momentum $k_i$, and $N$ is the number of occupied bands 
(counting only one of the Kramers degenerate pairs). 
Since Ti$_3$N$_2$F$_2$ possesses a hexagonal lattice, the time reversal invariant momenta $k_i$ are 
the $\Gamma$ point at \textbf{k}=(0,0,0) and the three $M$ points at \textbf{k}=(0,0.5,0), (0.5,0,0), and (0.5,0.5,0). 
Investigating the parities of the occupied bands at the 
$\Gamma$ and $M$ points, we find that $\delta(\Gamma)=-1$ and $\delta(M)=1$, and thus the 
$\mathbb{Z}_2$ invariant $\nu$ is 1. Therefore, we can conclude that Ti$_3$N$_2$F$_2$ is a 2D TI with a 
band gap as large as 0.05 eV within the HSE06 calculation.

The band inversion plays an important role in the nontrivial band topology. As we will discuss 
in the following, 
the band inversion in Ti$_3$N$_2$F$_2$ is not ascribed to the SOC. 
To shed light on it, the projected band structures onto $d$ orbitals of different Ti atoms 
without the SOC is shown in Fig.~\ref{bands2}. Here, we only consider $d$ orbitals of Ti atoms because the bands 
near the Fermi energy are composed mostly of these $d$ orbitals. 
In Ti$_3$N$_2$F$_2$, two inequivalent Ti atoms have different local symmetries. The symmetry of the inner 
Ti atom is $D_{3d}$, while the symmetry of the two outer Ti atoms is $C_{3v}$. 
According to the point symmetry of the crystal lattice, five $d$ orbitals of Ti atoms are split. 
As shown in Fig~\ref{bands2}, different combinations of 
$d$ orbitals lead to different parities. For example, at the $\Gamma$ point, the band labeled with 
$A_{1g}$ in Fig.~\ref{bands}(a) stems from the $d_{z^2}$ orbitals of inner Ti atom and outer ones. 
Due to the inversion symmetry, the $d_{z^2}$ orbitals of the outer Ti atoms are combined symmetrically to 
form a band with even parity. 
The band with $A_{2u}$ symmetry at the $\Gamma$ point possesses odd parity. Therefore, 
the $d$ orbitals of the inner Ti atom with $D_{3d}$ symmetry does not contribute to this state because the inversion center is located at the inner Ti atom and the $d$ orbitals have only the even parity. 
It is merely the antisymmetric combination of the $d_{z^2}$ orbitals of the outer Ti atoms that contributes 
to this state. On the other hand, $d_{xy}$ and $d_{x^2-y^2}$ orbitals of the three Ti atoms result into two-fold degenerate $E_g$ bands at the $\Gamma$ point near the Fermi energy [see Figs.~\ref{bands}(a) and \ref{bands2}]. As discussed in the following, the band inversion occurs at the $\Gamma$ point 
among the bands with $A_{2u}$ and $E_g$ symmetries, leading to $\delta(\Gamma)=-1$. 
In contrast, there is no band inversion at the $M$ point. 
As shown in Fig.~\ref{bands}(b), the SOC is important only to open the gap at the $\Gamma$ point. 
The similar band inversion mechanism was found in W$_2$CO$_2$~\cite{Hongming Weng4} and 
double transition metals M$'_2$M$''$C$_2$ 
(M$'$= Mo, W; M$''$= Ti, Zr, Hf) MXenes~\cite{Mohammad Khazaei2016}.



It is known that the order of energy bands may change upon applying strain to convert TIs into trivial insulators 
and vice versa~\cite{Yong Xu,Chen Si,Chen Si2}.  Next, we calculate the energy bands of Ti$_3$N$_2$F$_2$ 
in the presence of compressive and stretching strain. For simplicity, we assume that the space group does not 
change with the strain, but the corresponding atomic structures are fully optimized. 
Figure~\ref{strain}(a) shows the results when the lattice constant is compressed by 12\%. 
We find that the band with A$_{2u}$ symmetry at the $\Gamma$ point shift downwards in energies upon
increasing the compressive strain, and eventually it becomes occupied and locates below the bands with E$_g$ 
symmetry at $\sim10$\% of the compressive strain [see Fig.\ref{strain}(c)]. This implies that the topological property of the energy bands changes greatly because $\delta(\Gamma)=1$.

As shown in Fig.~\ref{strain}(b), when the lattice is stretched about 12\%, 
we find a Dirac-like dispersion with the Dirac point at the Fermi energy along the $\Sigma$ axis 
(the axis connecting the $\Gamma$ and $M$ points). 
This Dirac point is due to the accidental degeneracy~\cite{H. Fashandi} and indeed the SOC opens the gap 
as large as 0.02 eV [see Fig.~\ref{strain}(d)]. The point group along the $\Sigma$ axis is $C_s$, whose symmetry 
is lower than the symmetry at the $\Gamma$ point. According to the compatibility relation, the doubly degenerate 
bands with $E_g$ symmetry at the $\Gamma$ point is split into non-degenerate bands with $A'$ and $A''$ 
symmetries along the $\Sigma$ axis, where the band with $A''$ symmetry is higher in energy than that with $A'$ 
symmetry, as shown in Fig.~\ref{strain}(b). Note also that the energy band with $A_{1g}$ symmetry at the $\Gamma$ point becomes the band with $A'$ symmetry along the $\Sigma$ axis. The stretching of the lattice moves the band with $A_{1g}$ symmetry upwards and, around 5\% stretching, its energy becomes higher than that of the bands with $E_{g}$ symmetry, as shown in Fig.~\ref{strain}(c). 
At the same time, the $A'$ band, originating from the $A_{1g}$ band at the $\Gamma$ point, crosses the $A''$ band, branched off from the $E_g$ bands at the $\Gamma$ point, along the $\Sigma$ axis, forming the Dirac point at the Fermi energy. As shown in Fig.~\ref{strain}(d), the SOC lifts this accidental degeneracy by opening the gap. Since the $A_{1g}$ and $E_{g}$ bands have the same parity, the exchange of their ordering does not change the topological property, and Ti$_3$N$_2$F$_2$ is still a TI.

Finally, the edge states of Ti$_3$N$_2$F$_2$ nanoribbon are studied. The nanoribbon is made by cutting it 
along the \emph{a} axis [see Fig.~\ref{crystructure}(a)]. Figure~\ref{edge} summaries the results of the 
electronic band structures. It is clearly observed in Fig.~\ref{edge} that the emergent edge states 
cross the Fermi energy five times (i.e., odd number of times), revealing a non-trivial topology of the state.

\subsection{Electronic Properties of Zr$_3$N$_2$F$_2$ and Hf$_3$N$_2$F$_2$}

Since Zr and Hf atoms have very similar physical and chemical properties with Ti atom, next we study 
the electronic structures of Zr$_3$N$_2$F$_2$ and Hf$_3$N$_2$F$_2$, and their responses 
to the applied strain.

The energy bands of Zr$_3$N$_2$F$_2$ without and with the SOC are shown in Figs.~\ref{Zr_band}(a) and 
\ref{Zr_band}(b), respectively. 
Zr$_3$N$_2$F$_2$ is semimetal with having compensated electron and hole Fermi pockets. 
However, since the topmost and bottommost bands around the Fermi energy do not cross their bands, 
we can still calculate the $\mathbb{Z}_2$ invariant $\nu$ for bands denoted by black solid lines in 
Fig.~\ref{Zr_band}(b), and find that $\nu$ is 1. 
Therefore, Zr$_3$N$_2$F$_2$ has the same band topology as Ti$_3$N$_2$F$_2$. 
If the lattice is compressed as large as 3\%, indicated by the red arrow in Fig.~\ref{Zr_band}(e), 
the band with $A_{2u}$ symmetry at the $\Gamma$ point becomes lower in energy than the bands with $E_g$ 
symmetry [see Fig. \ref{Zr_band}(c) for the energy band under 5\% compression]. 
As a result, the band topology changes because $\delta(\Gamma)$ is now $+1$. 

On the other hand, if it is stretched, the band with $A_{1g}$ symmetry at the $\Gamma$ point moves upward in energy, whereas the bands with $E_g$ symmetry moves downward, and the order of these bands is eventually exchanged when the stretch strain exceeds 4\% [indicated by the blue arrow in Fig.~\ref{Zr_band}(e)]. At the same time, the Dirac-like dispersion appears along the $\Sigma$ axis. For example, the band structure with 7\% stretching is shown in Fig.~\ref{Zr_band}(d). This is analogous to the case of Ti$_3$N$_2$F$_2$ discussed above. The accidental degeneracy of the Dirac point is lifted by introducing the SOC and thus 
the stretched Zr$_3$N$_2$F$_2$ becomes a TI. For instance, in the case of 7\% stretching, the SOC opens the band gap as large as 0.05 eV [see Fig.~\ref{Zr_band}(f)].

The energy band of Hf$_3$N$_2$F$_2$ is shown in Fig.~\ref{Hf_band}. As opposed to Ti$_3$N$_2$F$_2$, we find that Hf$_3$N$_2$F$_2$ is a trivial insulator because the band with $A_{2u}$ symmetry at the $\Gamma$ point is occupied, while the bands with $E_g$ symmetry is unoccupied [see Fig.~\ref{Hf_band}(a)].  However, by applying the stretching strain as large as 5\%, the band inversion occurs among the bands with $A_{2u}$ and $E_g$ symmetries. As shown in Fig.~\ref{Hf_band}(b), the band with $A_{2u}$ symmetry shifts upwards and becomes unoccupied, whereas the bands with $E_g$ symmetry are now the valence bands. 
Therefore, the stretched Hf$_3$N$_2$F$_2$ has a nontrivial band topology. The band gap opens in the presence of SOC and can be as large as 0.1 eV within the HSE06 calculations with 5\% of the stretching strain [see Fig.~\ref{Hf_band}(c)]. 

\subsection{The stabilities with the tensile strains}

To confirm the stabilities of Ti$_3$N$_2$F$_2$, Zr$_3$N$_2$F$_2$ and Hf$_3$N$_2$F$_2$ in the presence of different tensile strains, it is straightforward to calculate their phonon spectra, respectively. The calculated phonon spectra are shown in Fig.~\ref{stability}. We find that these compressed and stretched structures can be stable in the range as we have discussed previously, because they have all positive frequencies. Our results are consistent with the measurements that MXene flakes are highly flexible materials with excellent tensile and compressive strengths~\cite{M. R. Lukatskaya, Zheng Ling}. However, for Ti$_3$N$_2$F$_2$, we plotted the result with 8.5\% stretching in Fig.~\ref{stability}(d), and one phonon branch, as indicated by the arrow, underwent a drop to the low frequency at \emph{K} point. By further stretching, Ti$_3$N$_2$F$_2$ is not stable because the vibrational branch with imaginary frequencies can be found. 

The modulation of the tensile strain in experiments is still a challenge. Different methods and techniques have been proposed to realize it. For example, the tensile tests were preformed by gluing MXene films onto supporting paper frames~\cite{Zheng Ling}. It is reported that by introducing polyvinyl alcohol (PVA) films, the tensile strength was improved significantly~\cite{Zheng Ling}. However, the experimental observation of our prediction with different tensile strains is still an open question.

\section{Conclusions}\label{sec:conclusion}

Previously, we have proposed that the functionalized transition metal carbides M$_2$CO$_2$ (M=W, Mo, and Cr) and M$'_2$M$''$C$_2$ (M$'$= Mo, W; M$''$= Ti, Zr, Hf) are 2D TIs~\cite{Hongming Weng4,Mohammad Khazaei2016}. Based on the first-principles calculations, here we have found that the topological property can be extended to functionalized nitride MXenes, such as Ti$_3$N$_2$F$_2$ and Zr$_3$N$_2$F$_2$. Although Hf$_3$N$_2$F$_2$ is a trivial insulator, it can be converted into a TI by strain. Their response to the strain indicates that these materials are promising candidates as future nano devices and strain sensors. Because of the easier production process by a selective chemical etching method, hosting QSHE at ambient condition, the high stability and antioxidant upon exposure to air, and consisting of environmental friendly elements, we expect that our finding will stimulate further experimental studies.

\section{Acknowledgments}
Y.L. would like to express his sincere thanks to the crew in the Center for Computational Materials 
Science of the Institute for Materials Research, Tohoku University, for their continuous support of the SR16000 
supercomputing facilities. M.K and A.R are grateful to RIKEN Advanced Center for Computing and Communication (ACCC) for the allocation of computational resource of the RIKEN supercomputer system (HOKUSAI GreatWave). Part of the calculations in this study were performed on Numerical Materials Simulator at NIMS. M.K. gratefully acknowledges the support by Grant-in-Aid for Scientific Research (No. 17K14804) from MEXT Japan.

\newpage

\begin{flushleft}
{\large \textbf{Captions} }
\end{flushleft}

\begin{figure}[tbp]
\includegraphics[width=1\textwidth]{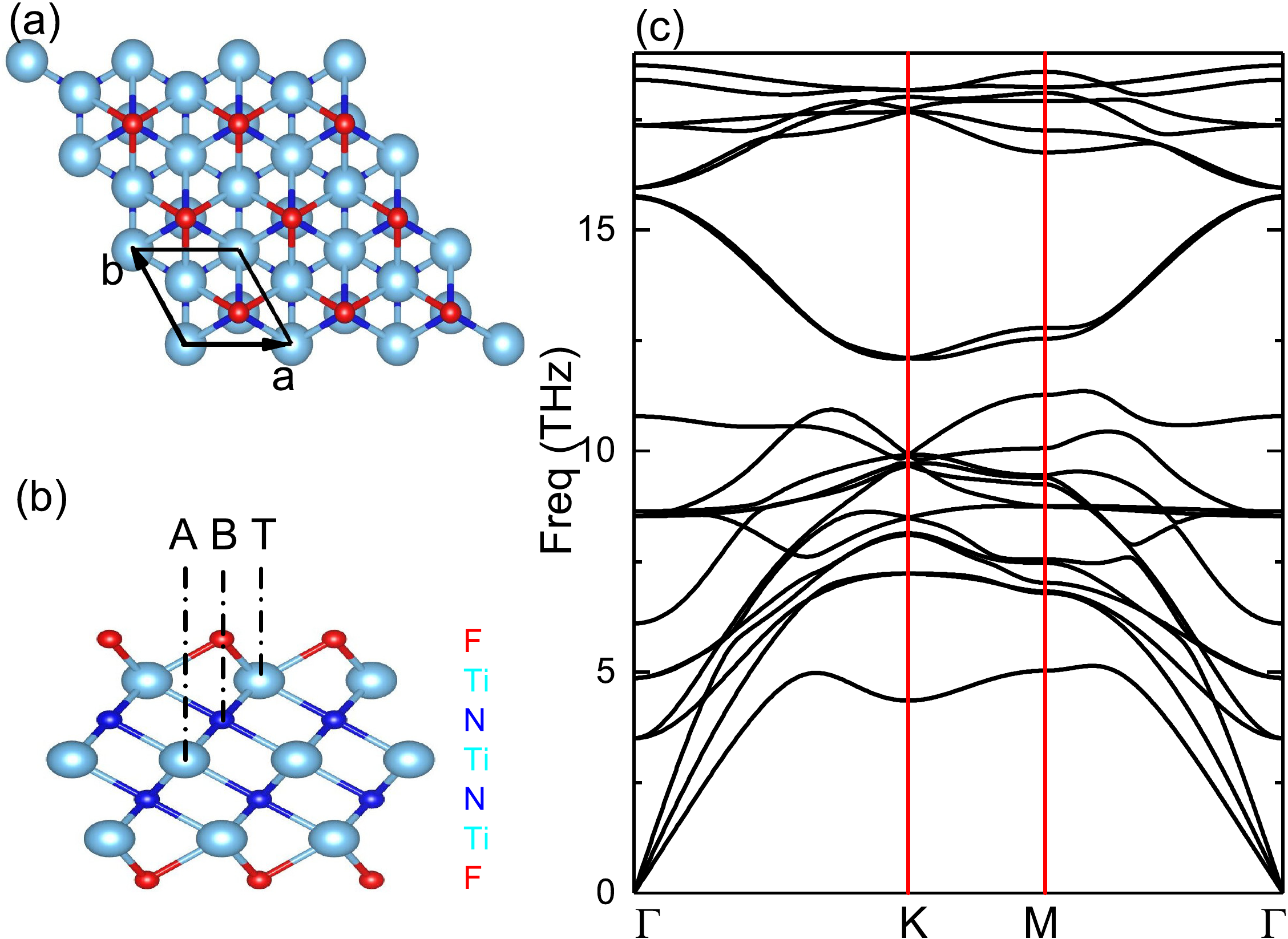}
\caption{Y. Liang \emph{et. al.}} \label{crystructure}
\end{figure}

\begin{figure}[tbp]
\includegraphics[width=1\textwidth]{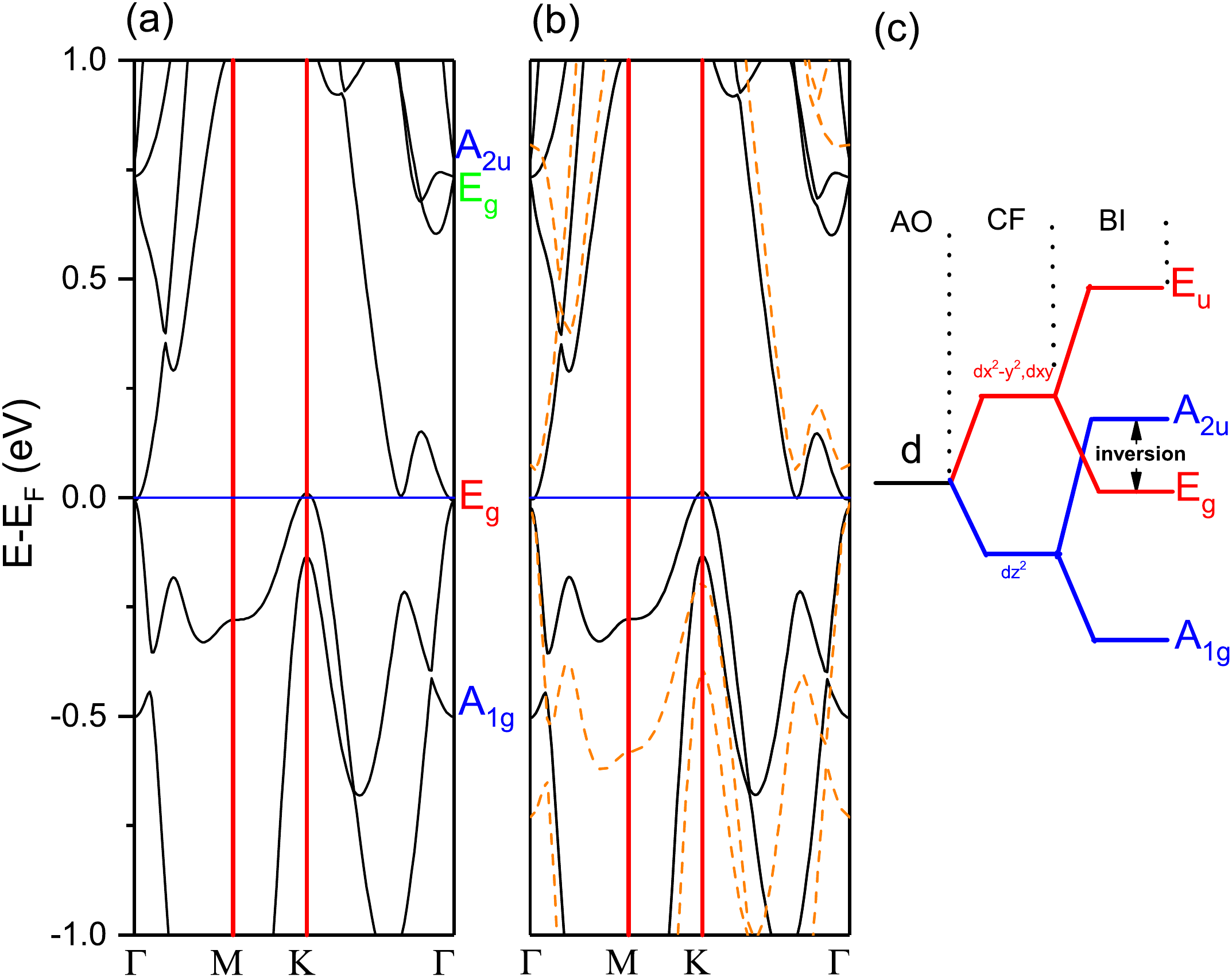}
\caption{Y. Liang \emph{et. al.}} \label{bands}
\end{figure}

\begin{figure}[tbp]
\includegraphics[width=1\textwidth]{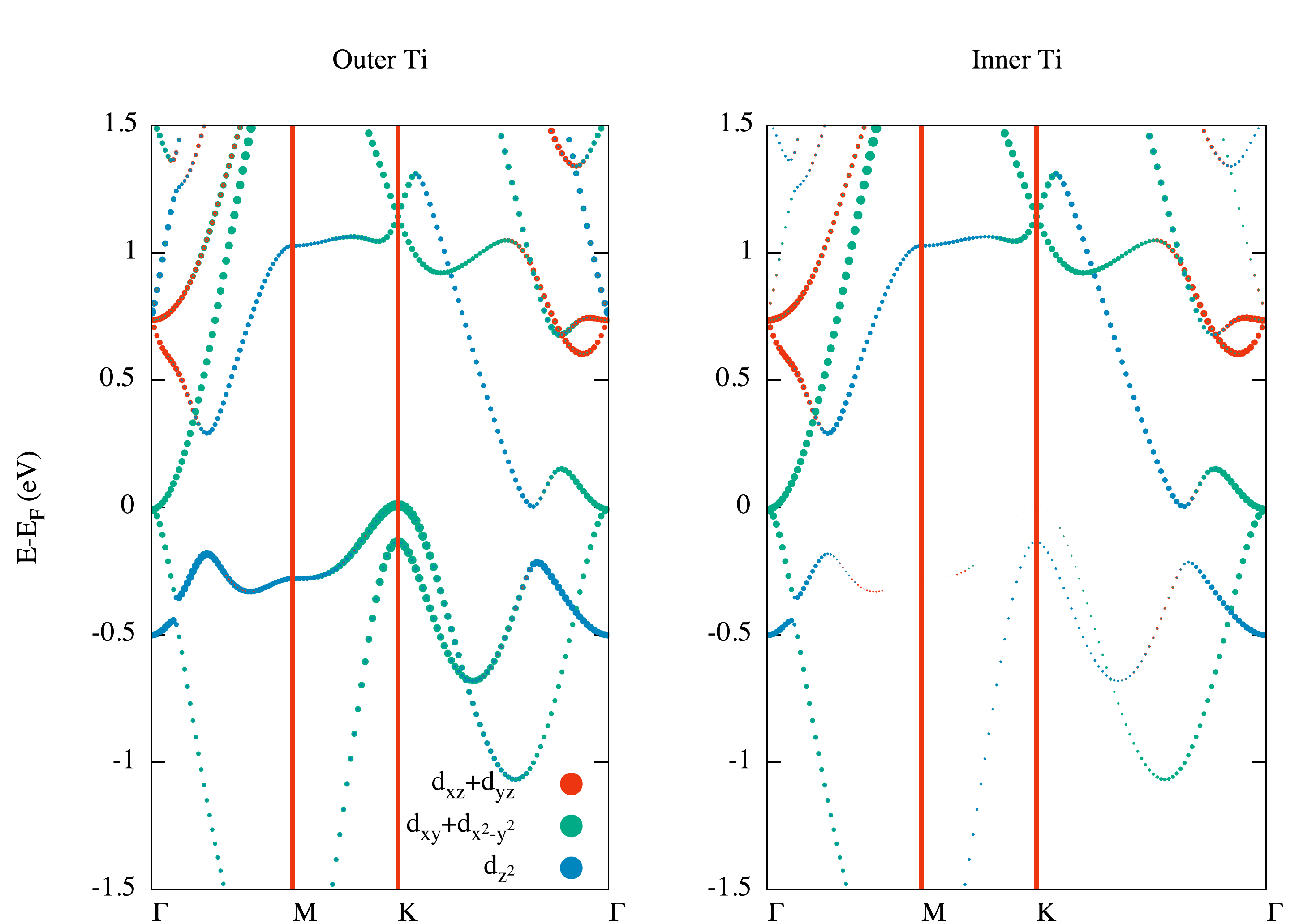}
\caption{Y. Liang \emph{et. al.}} \label{bands2}
\end{figure}

\begin{figure}[tbp]
\includegraphics[width=1\textwidth]{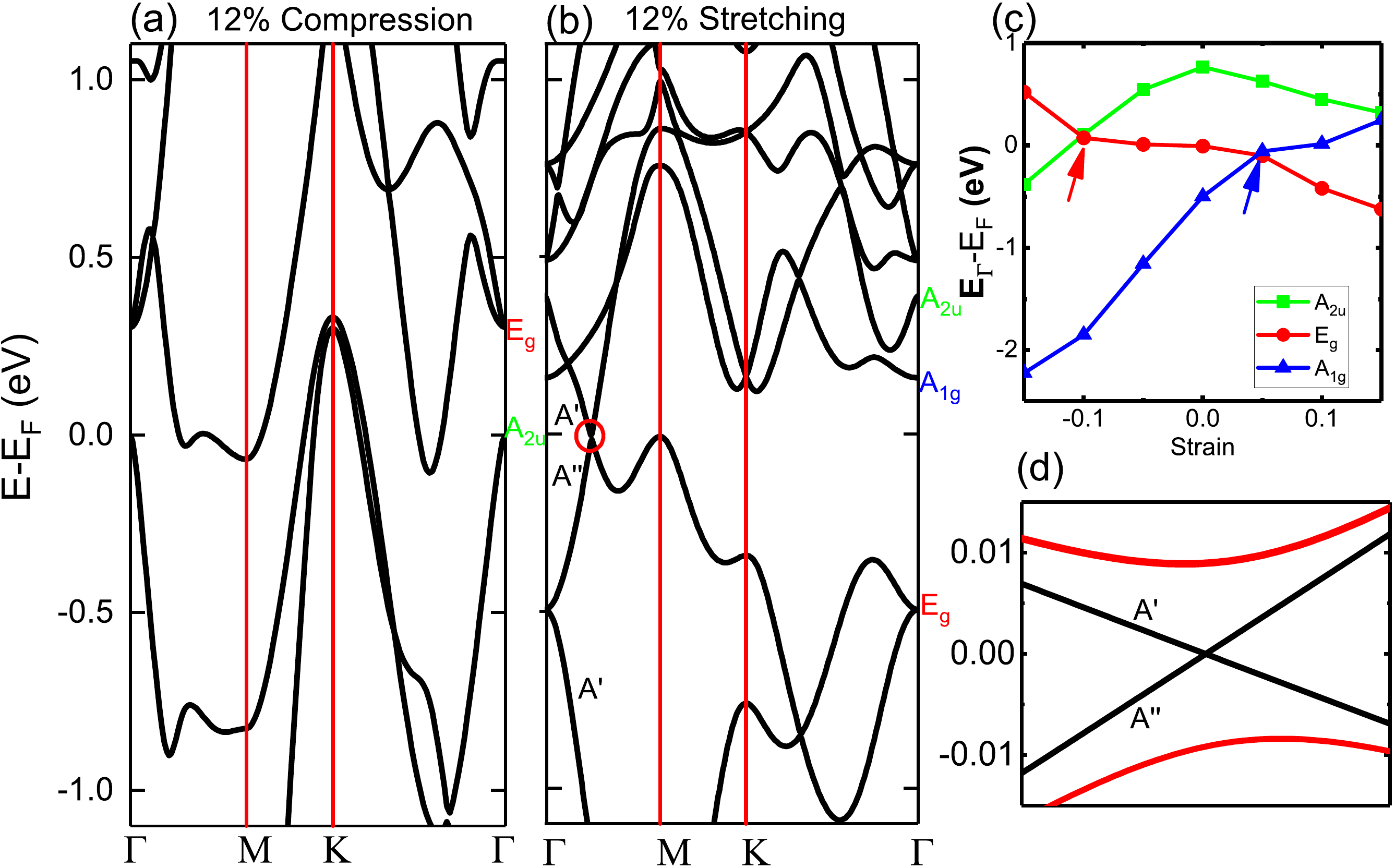}
\caption{Y. Liang \emph{et. al.}} \label{strain}
\end{figure}

\begin{figure}[tbp]
\includegraphics[width=1\textwidth]{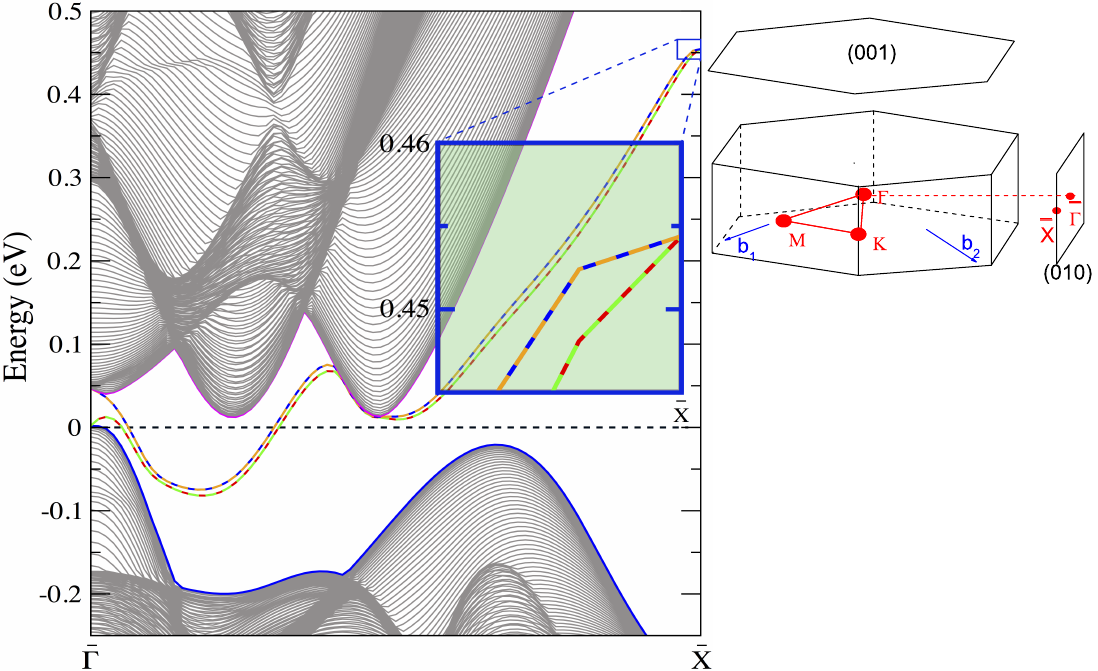}
\caption{Y. Liang \emph{et. al.}} \label{edge}
\end{figure}

\begin{figure}[tbp]
\includegraphics[width=1\textwidth]{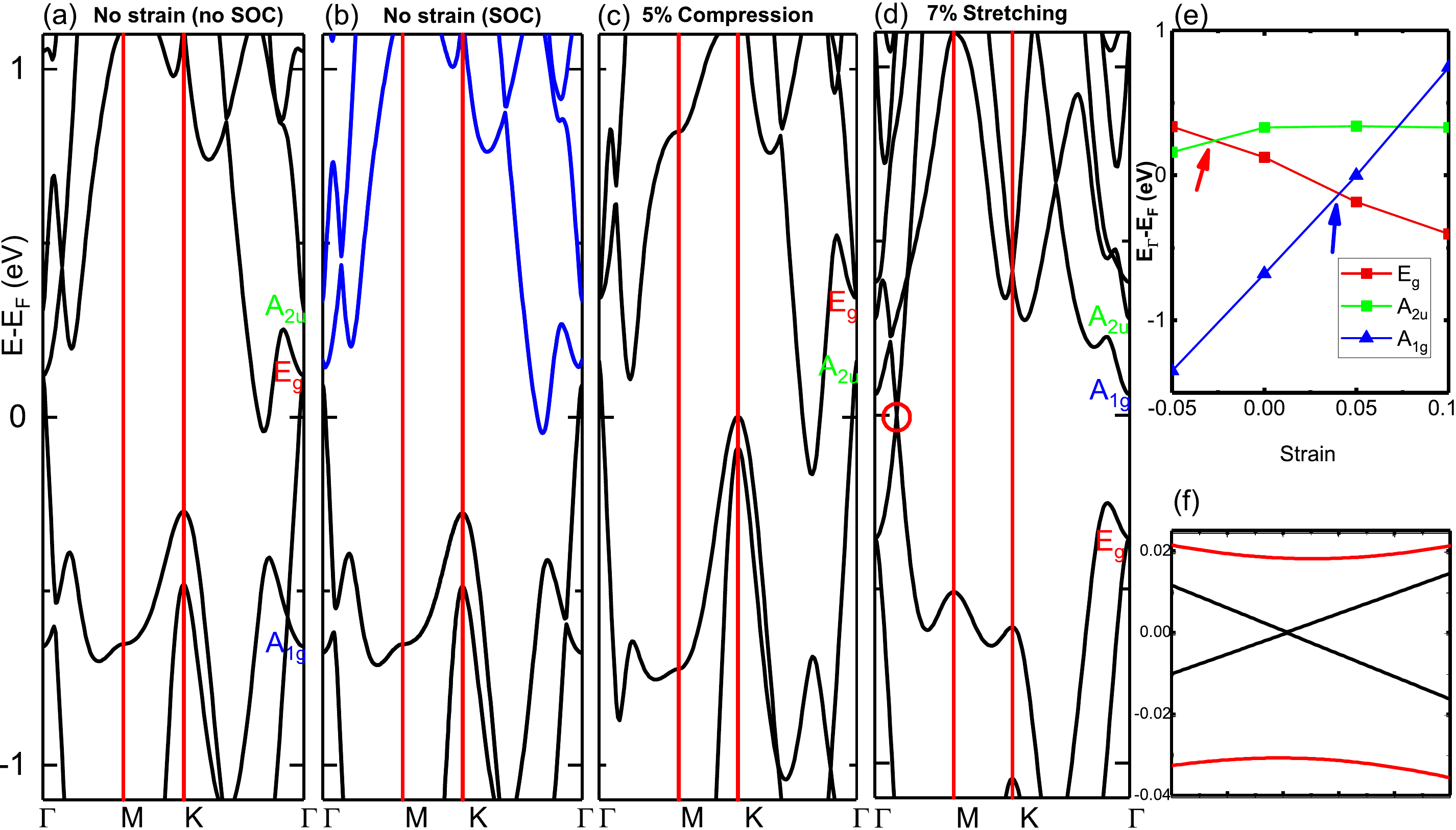}
\caption{Y. Liang \emph{et. al.}} \label{Zr_band}
\end{figure}

\begin{figure}[tbp]
\includegraphics[width=1\textwidth]{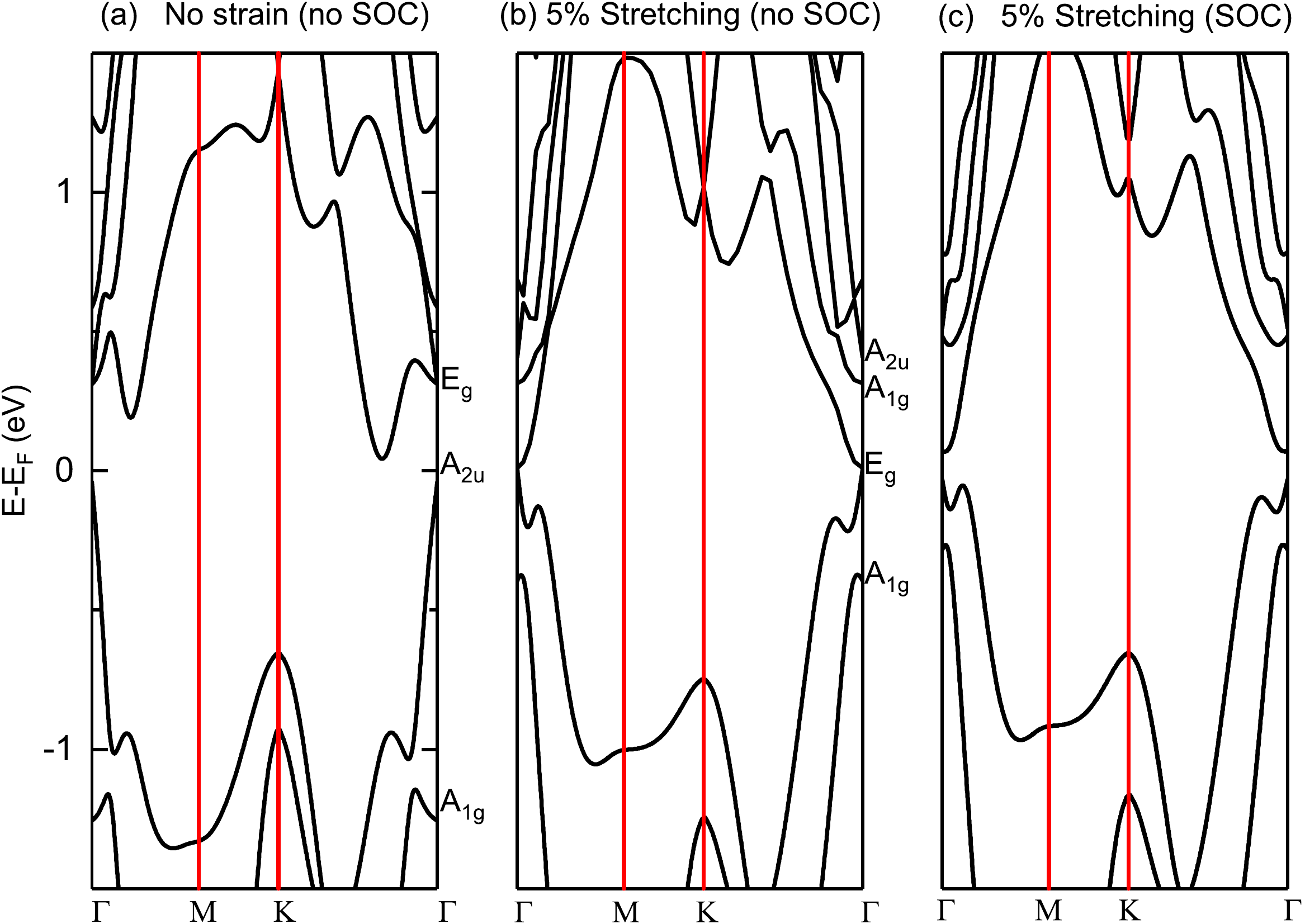}
\caption{Y. Liang \emph{et. al.}} \label{Hf_band}
\end{figure}

\begin{figure}[tbp]
\includegraphics[width=1\textwidth]{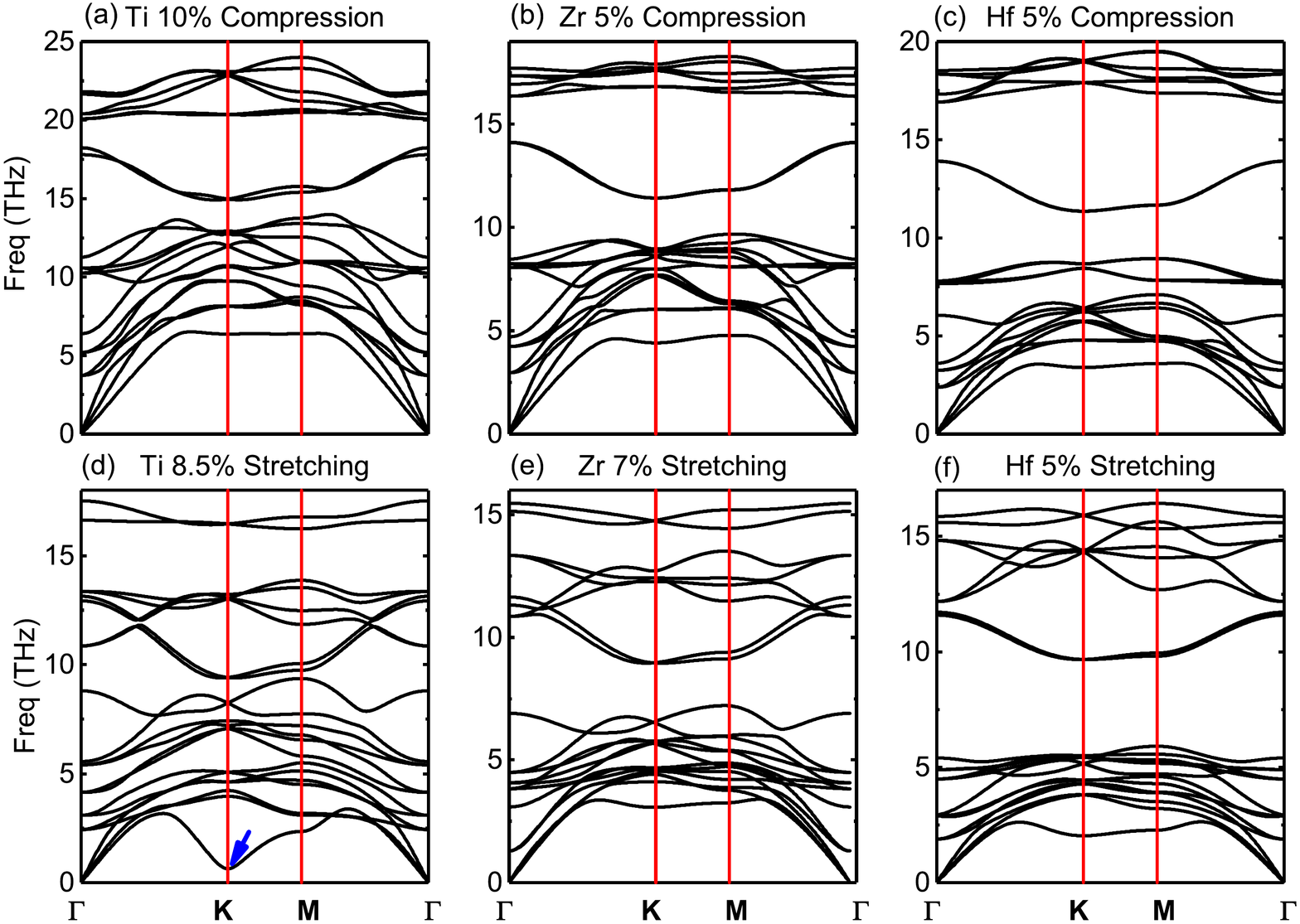}
\caption{Y. Liang \emph{et. al.}} \label{stability}
\end{figure}

\begin{description}

\item{FIG. \ref{crystructure}} (a) Top and (b) side views of optimized crystal structure of Ti$_3$N$_2$F$_2$. 
Here, Ti , N, and F atoms 
are represented by light blue, dark blue, and red spheres, respectively. Three possible positions, i.e., 
B (on top of N atom), T (on top of Ti atom in the outer layer), 
and A (on top of Ti atom in the central layer), decorating either side of the surfaces are shown in (b). 
The unit cell (black solid lines) with primitive lattice vectors $\vec a$ and $\vec b$ are indicated in (a). 
(c) The phonon dispersion spectrum for optimized Ti$_3$N$_2$F$_2$.

\item{FIG. \ref{bands}} The energy bands (solid lines) of Ti$_3$N$_2$F$_2$ calculated using the PBE functional 
(a) without and (b) with the SOC. The energy bands calculated using the hybrid HSE06 functional with the SOC 
are also shown by dashed lines in (b). 
$E_{\rm F}$ is the Fermi energy. The irreducible representations of the energy bands at the $\Gamma$ point 
close to the Fermi energy are indicated in (a). 
(c) A schematic figure of the band inversion mechanism in Ti$_3$N$_2$F$_2$. 
The $d$ type atomic orbitals (AOs) are split due to the crystal field (CF) of lattice with 
\emph{P$\overline{3}$m1} symmetry. Band inversion (BI) occurs between states with different parities. 

\item{FIG. \ref{bands2}} 
The projected energy bands onto $d$ orbitals of (a) two outer and (b) inner Ti atoms for Ti$_3$N$_2$F$_2$
calculated using the PBE functional without the SOC. $E_{\rm F}$ is the Fermi energy.   

\item{FIG. \ref{strain}} The energy bands of Ti$_3$N$_2$F$_2$ under the strain 
calculated using the PBE functional without the SOC. (a) The lattice is compressed by 12\%. (b) The lattice is stretched by 12\%. The red circle indicates the Dirac point in (b). The irreducible representations of the energy bands at the $\Gamma$ points and along the $\Sigma$ axis (the axis connecting the $\Gamma$ and $M$ points) 
near the Fermi energy $E_{\rm F}$ are indicated. (c) The energy evolution of three bands at the $\Gamma$ point with A$_{1g}$, A$_{2u}$, and E$_g$ symmetries, indicated in (a) and (b), when the lattice is compressed (negative strain) or stretched (positive strain). The red arrow indicates the compressive strain above which the band with A$_{2u}$ symmetry becomes lower in energy than that with E$_g$ symmetry, and thus the system becomes 
topologically trivial. The blue arrow implies the stretching strain above which the accidental degeneracy along the $\Sigma$ axis occurs. (d) The enlarged energy bands indicated by the red circles in (b). The accidental degeneracy at the Dirac point is lifted with the introduction of the SOC (red lines). 

\item{FIG. \ref{edge}} 
The edge states of Ti$_3$N$_2$F$_2$ cut along the \emph{a} axis [see Fig.~\ref{crystructure}(a)]. 
The Fermi energy is located at zero energy. The inset shows the enlarged energy bands around the 
$\bar X$ point. It is clearly observed that the edge states cross the Fermi energy five times and 
are doubly degenerate at the $\bar X$ point. 

\item{FIG. \ref{Zr_band}} The energy bands of Zr$_3$N$_2$F$_2$ with no strain (a) in the absence of 
SOC and (b) in the presence of  SOC, 
and with (c) 5\% compression and (d) 7\% stretching in the absence of SOC. 
The irreducible representations of the energy bands at the $\Gamma$ point near the Fermi energy $E_{\rm F}$ are indicated. (e) The energy evolution of the bands at the $\Gamma$ point with $A_{2u}$, $A_{1g}$, and $E_g$ symmetries. The red and blue arrows imply the occurrence of bands crossing with varying the strain. (f) The enlarged energy bands (black lines) indicated by the red circle in (d). The accidental degeneracy at the Dirac point is lifted with the SOC (red lines). All calculations are performed using the PBE functional.


\item{FIG. \ref{Hf_band}} The energy bands of Hf$_3$N$_2$F$_2$ calculated using the HSE06 hybrid 
functional (a) without any strain and the SOC, (b) with 5\% stretching strain but no SOC, and (c) with 
5\% stretching strain and the SOC. The irreducible representations of the energy bands at the $\Gamma$ point near the Fermi energy $E_{\rm F}$ are indicated. 

\item{FIG. \ref{stability}} The phonon spectra of Ti$_3$N$_2$F$_2$, Zr$_3$N$_2$F$_2$ and Hf$_3$N$_2$F$_2$ with different tensile strains. 

\end{description}

\clearpage\newpage

\clearpage\newpage

\clearpage\newpage

\clearpage\newpage

\clearpage\newpage

\clearpage\newpage

\clearpage\newpage

\clearpage\newpage

\end{document}